\documentstyle[sprocl]{article}

\input{epsf}

\def\Journal#1#2#3#4{{#1} {\bf #2}, #3 (#4)}

\def\PREP{\em Phys. Rep.}

\def\NPB{{\em Nucl. Phys.} B}
\def\PLB{{\em Phys. Lett.}  B}

\def\PRD{{\em Phys. Rev.} D}

\def\d{\partial}

\def\eps{\epsilon}

\def\as{\alpha_S}
\def\pts{(\phi^3)_6}

\def\nn{\nonumber}
\def\dk{\f{d^6k}{(2\pi)^6}}
\def\kh{{\hat k}}

\newcommand{\f}[2]{\frac{#1}{#2}}

\def\be{\begin{equation}}
\def\ee{\end{equation}}
\def\bea{\begin{eqnarray}}
\def\eea{\end{eqnarray}}
\begin{document}

\title{FRACTURE FUNCTIONS: FACTORIZATION AND EVOLUTION $^{a,b}$}

\author{M. GRAZZINI}

\address{INFN, Gruppo Collegato di Parma,\\43100 Parma, Italy\\and\\
Theory Division CERN, Geneve 23,\\CH 1211 Switzerland}


\maketitle\abstracts{Fracture functions and their evolution equation
are reviewed.
%in the framework of cut vertices and Jet Calculus.
Some phenomenological applications are briefly discussed.}

%\section{Introduction}
\footnotetext{$^a$Talk given at the DIS98 Workshop, Brussels, Belgium, April 4-8, 1998}
\footnotetext{$^b$Work done in collaboration with G. Camici, L. Trentadue and G. Veneziano}
Fracture functions \cite{tv} have been introduced to extend the usual QCD improved parton picture
of semi-inclusive deep inelastic processes to the low transverse momentum region of phase space, where the target fragmentation contribution becomes important.
Trentadue and Veneziano \cite{tv} proposed to describe such
contribution as a convolution of a new phenomenological distribution, the fracture function, with
a hard cross section
\be
\label{st}
\sigma_T=\int \f{dx^\prime}{x^\prime}
M^i_{AA^\prime}(x^\prime,z,Q^2)
{\hat \sigma}_i(x/x^\prime,Q^2).
\ee
The fracture function
$M^i_{AA^\prime}(x,z,Q^2)$ represents the
probability of finding the parton
 $i$ in the
hadron $A$ with momentum fraction $x$ while observing the hadron $A^\prime$ in the inclusive final state with momentum fraction $z$.
In the case in which the momentum transfer $t=|(p_A-p_{A^\prime})^2|$ is measured we define \cite{cut}
${\cal M}^i_{AA^\prime}(x,z,t,Q^2)$, a $t$-dependent (extended)
fracture function.

The same idea of fracture functions implies the existence of a new
factorization theorem which allows to write eq. (\ref{st}).
In the case of inclusive DIS one can use OPE
%to get a prediction for the forward scattering amplitude which,
%by using a dispersion relation can be turned into a factorized formula for the
%inclusive cross section,
but in semi-inclusive processes
the straightforward application of OPE fails.
The problem comes from the fact that OPE gives a prediction for amplitudes,
whereas
we need an expansion for cross-sections, i.e. cut amplitudes.
A possible way out is to use cut vertices.
%\section{Cut vertices}

The cut-vertex expansion \cite {mue}
is a generalization of OPE where local operators
are replaced by non-local objects, i.e. cut vertices.
Such an expansion allows to treat more general processes.
%Let us consider DIS in $\pts$ off the current $J=1/2 \phi^2$
%\begin{equation}
%F(p,q)=\sum_\tau \int V_\lambda(p,k) H_\tau(\kh,q) \dk\nn
%\end{equation}
%where $\kh=(k_+,{\bf 0},0)$.
%Define
%\bea
%v_\l(p^2,x)& = &\int V_\l(p,k) x \delta\left(x-\f{k_+}{p_+}\right)\dk\nn\\
%C_\tau(x,Q^2)& = &\hspace{1mm} H_\tau(k^2=0,x,q^2)\nn
%\eea
%In the large $Q^2$ we can write for the moments of the inclusive cross section
%\be
%\label{fact1}
%W_\sigma(p,q)\simeq v_\sigma(p^2) C_\sigma(Q^2)
%\ee
%Where v
Let us consider semi-inclusive DIS in $\pts$,
that is the
%process
scattering reaction
$p+J(q)\to p^\prime+X$ where $J=1/2~\phi^2$. In the region $p^\prime_t\ll Q$ or equivalently $t\ll Q^2$,
the relevant diagrams are those in Fig.\ref{decomposition2}:
%%====================================
\begin{figure}[htb]
\begin{center}
\begin{tabular}{c}
\epsfxsize=5.5truecm
\epsffile{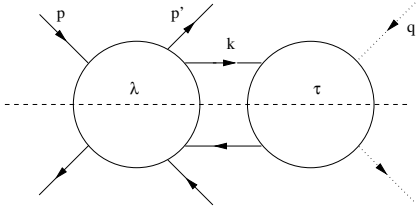}\\
\end{tabular}
\end{center}
\caption{Relevant decomposition in the region $t\ll Q^2$}
\label{decomposition2}
\end{figure}
%%===================================
\begin{equation}
W(p,p^\prime,q)=\sum_\tau \int T_\lambda(p,p^\prime,k) H_\tau(\kh,q) \dk\nn
\end{equation}
where, given a vector $k$, $\kh=(k_+,{\bf 0},0)$.
Define:
\be
\label{cu}
v_\lambda(p,p^\prime,\bar{x})=\int T_\lambda(p,p^\prime,k)\bar{x}
\delta\left(\bar{x}-\f{k_+}{p_+-p^\prime_+}\right)\f{d^6k}{(2\pi)^6}
\ee
\be
\label{co}
C_\tau(x,Q^2)=H_\tau(k^2=0,x,q^2)\nn
\ee
where $\bar{x}=x/(1-z)$. By taking moments we can write
\be
\label{fact2}
W_\sigma(p,p^\prime,q)\simeq \sum_\tau v^\sigma_\lambda(p,p^\prime)
C_\tau^\sigma(Q^2)
\equiv v_\sigma(p,p^\prime) C_\sigma(Q^2).
\ee
Here $v_\sigma(p,p^\prime)$ is a new cut vertex which contains the long
distance dependence of the cross section, whereas $C_\sigma(Q^2)$ is a coefficient
function which is calculable as usual in perturbation theory.
The expansion is technically obtained \cite{paper} by
constructing an identity so as to isolate
the leading term from the remainder, and care has to be taken to remove the UV divergences
%which plague
hidden in
eqs.(\ref{cu}) and (\ref{co}).
Eventually one has to prove that the leading term
is {\em really} leading.
This is not obvious since there is no Weinberg theorem
for cut amplitudes. Nevertheless,
in order to find the leading behavior in the large $Q^2$ limit one can look at
the singularities \cite{sterman} at $p^2$,$~p^{\prime 2}$,$~t\to 0$.
We find that the leading singularities are given by diagrams of the kind of Fig.\ref{decomposition2}, and so we can say that the cut-vertex expansion really gives the leading contribution.
%%====================================
%\begin{figure}[htb]
%\begin{center}
%\begin{tabular}{c}
%\epsfxsize=7truecm
%\epsffile{leading.eps}\\
%\end{tabular}
%\end{center}
%\caption{{Deep inelastic structure function in $\pts$}}
%\end{figure}
%%====================================
In QCD one has the complication that soft gluon contributions are
not suppressed as in $\pts$ by power counting.
However, by using gauge invariance,
it can been shown that they cancel out \cite{collins}.
%so we can say that the factorization theorem is now proved.
%\section{Evolution pattern}

The coefficient function appearing in the cut-vertex expansion
is exactly the same as in the inclusive case since it comes from the hard part of the
graphs. This means that
the $Q^2$ evolution is dictated by the anomalous dimension of the same
minimal twist local operator. By using renormalization group
we can write eq. (\ref{fact2}) in QCD as
\be
W_n(z,t,Q^2)=\sum_i {\cal M}^i_n(z,t,Q^2)~C_n^i (1,\as(Q^2))
\ee
%%%%%%%%%%%%%%%%%%%%%%%%%%%%
% definizione dal cut vertex
%%%%%%%%%%%%%%%%%%%%%%%%%%%%
where we have defined \cite{cut}
\begin{equation}
{\cal M}^j_n(z,t,Q^2)\equiv V^i_n(z,t,Q_0^2)
\left[ e^{\textstyle \int_{\as}^{\as(Q^2)} d\alpha
\f{\gamma^{(n)}(\alpha)}{\beta(\alpha)}}\right]_{ij}\nn
\end{equation}
just in terms of the cut vertex $V^i_n(z,t,Q_0^2)$.
It follows that
the evolution equation for $t$-dependent fracture function is a standard
DGLAP equation
\be
Q^2 \f{\d}{\d Q^2}{\cal M}^j_{A,A'}(x,z,t,Q^2)= \f{\as (Q^2)}{2\pi}
\int_{\f{x}{1-z}}^1 \f{du}{u} P_i^j(u)
{\cal M}^i_{A,A'}(x/u,z,t,Q^2).
\ee
In the perturbative region of $t$ we can give a definition \cite{evo} of
${\cal M}_{AA^\prime}^j(x,z,t,Q^2)$ based on Jet
Calculus \cite{jet}
%%====================================
\begin{figure}[htb]
\begin{center}
\begin{tabular}{c}
\epsfxsize=4.5truecm
\epsffile{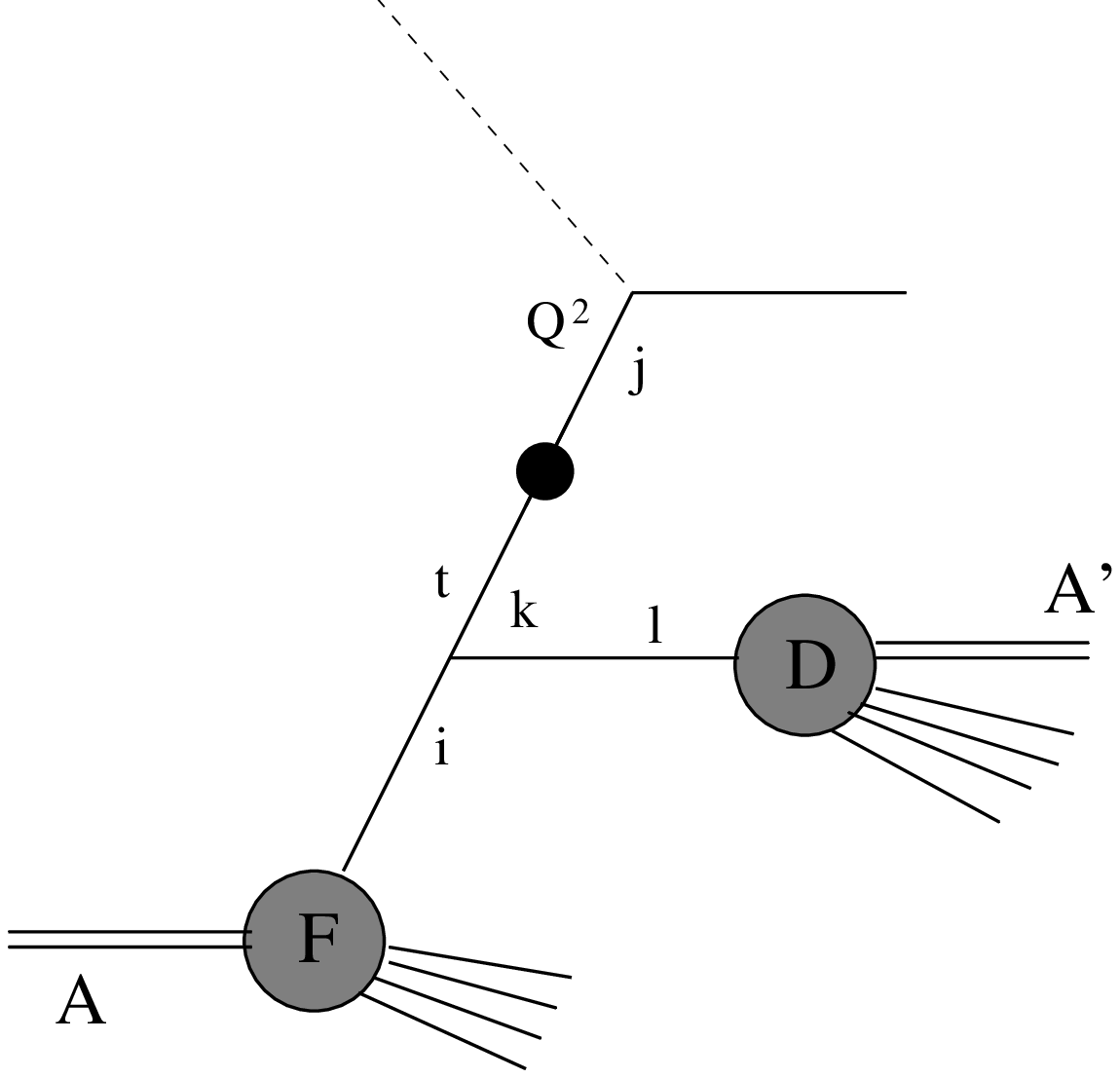}\\
\end{tabular}
\end{center}
\caption{{Perturbative definition of the $t$-dependent fracture function}}
\end{figure}
%%====================================
%%%%%%%%%%%%%%%%%%%%%%%%%%
% definizione perturbativa
%%%%%%%%%%%%%%%%%%%%%%%%%%
\bea
{\cal M}_{AA^\prime}^j(x,z,t,Q^2)&=&\f{\as (t)}{2\pi t} \int^{1-z}_x\f{dr}{r}\int^1_{z+r}\f{dw}{w(w-r)}
F_A^i(w,t){\hat P}_i^{kl}\left(\f{r}{w}\right)\times\nn\\
&\times& D_{l,A'} \left(\f{z}{w-r},t\right) E_k^j(x/r,t,Q^2)
\eea
where ${\hat P}_i^{kl}$ is the real splitting function and $E_k^j(x,t,Q^2)$
is the evolution kernel from the scale $t$ to $Q^2$.
If we define the ordinary fracture function as an integral up to a cut off
of order $Q^2$, say $\eps Q^2$, the inhomogeneous evolution equation \cite{tv}
for fracture functions is recovered
up to $\log \eps$ corrections
%%%%%%%%%%%%%%%%%%%%%%%%%%
% equazione inomogenea
%%%%%%%%%%%%%%%%%%%%%%%%%%
\bea
&Q^2& \f{\d}{\d Q^2}M^j_{A,A'}(x,z,Q^2)=
\f{\as(Q^2)}{2\pi}
\int^1_{\f{x}{1-z}}\f{du}{u} P_i^j(u) M^i_{A,A'}(x/u,z,Q^2)\nn\\
&+&\f{\as (Q^2)}{2\pi}\int^{\f{x}{x+z}}_x \f{du}{x(1-u)}
F_A^i(x/u,Q^2){\hat P}_i^{jl}(u)
D_{l,A'}\left(\f{zu}{x(1-u)},Q^2\right).
\eea
Fracture functions are now measured at HERA and the scaling violations
observed in experimental data are consistent with the evolution pattern presented here \cite{df}.

Fracture functions give the possibility of selecting interesting
channels.
%Observing semi-inclusive processes at large $z$ with particular
%choices of the observed hadron is almost equivalent to performing
%inclusive DIS on virtual hadronic targets.
This fact could be used to test target independence of suppression of
the first moment of the polarized proton structure function \cite{sv}.
Another interesting possibility could be
to select a gluon by requiring a proton in the inclusive final state and study
correlations with heavy quark production.

What about hadron-hadron scattering?
One would like to use HERA data on fracture functions to give predictions
for hadron-hadron scattering.
However the
general claim is that the factorization theorem for diffractive hadron-hadron scattering fails to hold, since
the cancellation at work in the inclusive case does not apply
here.
Nevertheless we believe that further work is needed to asses this conclusion. 

%\section{Conclusions}
Summing up, we can say that fracture functions and their evolution equations are now well
understood.
We have given them an interpretation based on a formalism which is a
direct generalization of OPE.
Having proved the factorization theorem,
fracture functions can now be reliably used to describe semi-inclusive
hard processes in the target fragmentation region.

%\section*{Acknowledgments}

\end{document}